\documentclass[aps,prl,superscriptaddress,reprint]{revtex4-1}
\usepackage{amsmath}
\usepackage{amssymb}
\usepackage{graphicx}
\usepackage[colorlinks,urlcolor=blue]{hyperref}
\usepackage{url}
\usepackage{color,xcolor}
\usepackage{ulem}
\usepackage{float}
\usepackage{epstopdf}
\begin{document}

\title{Theoretical study of the crystal and electronic properties of $\alpha$-RuI$_3$}
\author{Yang Zhang}
\author{Ling-Fang Lin}
\affiliation{Department of Physics and Astronomy, University of Tennessee, Knoxville, Tennessee 37996, USA}
\author{Adriana Moreo}
\author{Elbio Dagotto}
\affiliation{Department of Physics and Astronomy, University of Tennessee, Knoxville, Tennessee 37996, USA}
\affiliation{Materials Science and Technology Division, Oak Ridge National Laboratory, Oak Ridge, Tennessee 37831, USA}

\date{\today}

\begin{abstract}
The material $\alpha$-RuCl$_3$, with a two-dimensional Ru-honeycomb sublattice, has attracted considerable attention
because it may be a realization of the Kitaev quantum spin liquid (QSL). Recently, a new honeycomb material,
$\alpha$-RuI$_3$, was prepared under moderate high-pressure and it is stable under ambient conditions.
However, different from $\alpha$-RuCl$_3$, $\alpha$-RuI$_3$ was reported to be a paramagnetic metal without long-range magnetic order down to $0.35$ K. Here, the structural and electronic properties of the quasi-two-dimensional $\alpha$-RuI$_3$ are theoretically studied. First, based on first-principles density functional theory (DFT) calculations, the ABC stacking honeycomb-layer $R\overline{3}$ (No. 148) structure is found to be the most likely stacking order for $\alpha$-RuI$_3$ along the $c$-axis. Furthermore, both $R\overline{3}$ and $P\overline{3}1c$ are dynamically stable because no imaginary frequency modes were obtained in the phononic dispersion spectrum without Hubbard $U$. Moreover, the different physical behavior of $\alpha$-RuI$_3$ compared to $\alpha$-RuCl$_3$ can be understood naturally. The strong hybridization between Ru $4d$ and I $5p$ orbitals decreases the ``effective'' atomic Hubbard repulsion, leading the electrons of RuI$_3$ to be less localized than in RuCl$_3$. As a consequence, the effective electronic correlation is reduced from Cl to I, leading to the metallic nature of $\alpha$-RuI$_3$. Based on the DFT+$U$ ($U_{\rm eff} = 2$ eV), plus spin-orbital coupling (SOC), we obtained a spin-orbit Mott insulating behavior for $\alpha$-RuCl$_3$ and, by the same procedure, a metallic behavior for $\alpha$-RuI$_3$, in good agreement with  experimental results. Furthermore, when introducing a large (unrealistic) $U_{\rm eff} = 6$ eV, the spin-orbit Mott gap opens in $\alpha$-RuI$_3$ as well, supporting the physical picture we are proposing. Our results provide guidance to experimentalists and theorists working on two-dimensional transition metal triiodides layered materials.
\end{abstract}

\maketitle

\section{I. Introduction}
Due to their rich physical properties, low-dimensional materials continue attracting considerable attention in the Condensed Matter community~\cite{Bednorz:Cu,Dagotto:rmp94,Stewart:Rmp,Dagotto:rp,Dagotto:science05,Grioni:JPCM,Scalapino:Rmp,Dagotto:Rmp,Monceau:ap,Lin:prm19,Lin:prl19,miao:pnas2020,zhu:prb102,sen:2020,Zhang:prb21,Lin:prl21,mazza:2021,zhang:2021,lin:2021}. In systems with $3d$ transition-metal (TM) atoms, the electronic correlation couplings (i.e. Hubbard repulsion $U$ and Hund coupling $J_H$) play a key role to understand their physical properties. Their spin-orbital coupling (SOC) $\lambda$ is considered to be negligible. In those compounds, a wide variety of remarkable physical phenomena have been found driven by the bandwidth $W$ (corresponding to the kinetic hopping parameter $t$) and the electronic correlation couplings. The unusual states induced include high-T$_c$ superconductivity~\cite{Bednorz:Cu,Kamihara:Jacs,Dai:Np,Li:Nature,Zhang:prb20}, ferroelectricity triggered by spin or charge ordering~\cite{Brink:jpcm,Lin:prm17,Dong:nsr,Zhang:prb20-2}, orbital ordering~\cite{Tokura:science,Pandey:prb,Lin:prm21}, and charge or spin density waves~\cite{DW,Monceau:ap,Zhang:prbcdw}.

However, the $4d$ and $5d$ orbitals are more spatially extended than the $3d$ orbitals, leading to an increased hopping $t$ in the $4d/5d$ case. Furthermore, $U$ and $J_H$ are also reduced in the $4d/5d$ systems, compared to those for $3d$
electrons~\cite{JH:prb1,JH:prb2}. Moreover, the SOC parameter $\lambda$ is enhanced in $4d/5d$ systems~\cite{SOC}, inducing comparable values of $\lambda$ with $U$ and $J_H$. In this case, several intriguing electronic phases have been reported in $4d$ and $5d$ low-dimensional materials. In some dimer systems with $4d$ or $5d$ TM atoms, an interesting orbital-selective Peierls phase could be stable~\cite{Streltsovt:prb14,Zhang:ossp} when the intrahopping $t$ is larger than the typical Hund couplings. This phase resembles the previously discussed orbital-selective Mott phase~\cite{OSMP,Patel:osmp,Herbrych:osmp1,Herbrych:osmp2} but with the localized band induced by a Peierls distortion, instead of Hubbard interactions~\cite{Streltsovt:prb14,Zhang:ossp}. The Hubbard repulsion $U$ can lead to the localization of the spin-orbit coupled pseudospin degrees of freedom, resulting in a ``spin-orbit Mott'' insulating phase~\cite{Kim:prl08,Jackeli:prl08,Lu:am20}.

More interestingly, due to the strong bond-dependent anisotropic coupling among spins, a QSL ground state due to spin quantum fluctuations and frustration is theoretically obtained in the spin-$1/2$ honeycomb lattice via the Kitaev model~\cite{Kitaev:aop}. Honeycomb lattice materials with spin-$1/2$ were proposed to realize the Kitaev physics, such as the $5d^5$ iridates $A_2$IrO$_3$ ($A$ = Na, Li)~\cite{Chaloupka:prl08,Singh:prl12,Choi:prl12,Mazin:prl12,Rau:prl14,Kim:prx20}. In those systems, the concept of spin-$1/2$ arises from the effective $J_{\rm eff} = 1/2$ pseudospins induced by the strong SOC and crystal-field splitting~\cite{Kim:prl08}. However, due to substantial lattice distortions, such as dimerization under hydrostatic pressure, the $J_{\rm eff} = 1/2$ physical picture is destroyed~\cite{Bastien:prb18,Li:prm19}, and the Kitaev QSL is not realized.

A related Kitaev QSL candidate material is $\alpha$-RuCl$_3$ with a $4d^5$ electronic configuration analogue to the $5d^5$ iridates~\cite{Plumb:prb14}. This material also forms layered two-dimensional honeycomb structures, and the $4d^5$ electronic configuration of Ru is in a low-spin state with $S = 1/2$, producing $J_{\rm eff} = 1/2$ pseudospins~\cite{Johnson:prb15,Koitzsch:prb16}. At ambient conditions, $\alpha$-RuCl$_3$ exhibits spin-orbital Mott insulating behavior with a zigzag antiferromagnetic (AFM) ordering at $7-13$ K~\cite{Koitzsch:prb16,Cao:prb16}. Several stacking orders have been reported belonging to different space group, such as $C2/m$ (No. 12)~\cite{Johnson:prb15,Koitzsch:prb16,Banerjee:nm}, $P3_{1}12$ (No. 151)~\cite{Kubota:prb15}, and R$\overline{3}$ (No. 148)~\cite{Glamazda:prb17}.  Its unconventional interesting behavior, such as highly unusual magnetic excitations, the emergence of Majorana fermions and possible Kitaev QSL, attracted considerable attention both in experiments and theories related to this compound~\cite{Banerjee:nm,Glamazda:prb17,Kasahara:prl17,Kim:prb15,Hou:prb17,Eichstaedt:prb19,Baek:prl17,Zheng:prl17,Do:np17,Banerjee:science19,Banerjee:npjqm,Kasahara:nature}.

Very recently, a new honeycomb-structured material $\alpha$-RuI$_3$ has been synthesized at moderately high pressures~\cite{Ni:arXiv,Nawa:arXiv}. In general, considering the atomic number of I, the SOC effect should be larger than in Cl, which may lead to more interesting physical properties in RuI$_3$. Before the experimental preparation of $\alpha$-RuI$_3$, there were only a few theoretical studies focusing on the monolayer form~\cite{Huang:prb16,Ersan:jmmm19}. Preliminary characterization reveals metallic and paramagnetic behavior, with the absence of long-range magnetic order down to $0.35$ K~\cite{Ni:arXiv,Nawa:arXiv}. For $\alpha$-RuI$_3$ two different stacking orders were reported along the $c$-axis: the $R\overline{3}$ (No. 148) structure with a three-layer ABC stacking honeycomb-layer centrosymmetric rhombohedral symmetry~\cite{Ni:arXiv}, and a two-layered honeycomb structure model with the space group $P_{3}1c$ (No. 163)~\cite{Nawa:arXiv}. In each RuCl$_6$ plane, the honeycomb layers are built of edge-sharing RuI$_6$ octahedra. Different from the Ru-Cl bonds in $\alpha$-RuCl$_3$~\cite{Johnson:prb15}, the Ru-Ru bonds are identical with a Ru-Ru bond of length $3.92$~\AA~\cite{Ni:arXiv}. The van der Waals (vdW) layer distance is about $6.3$ \AA, larger than the value for $\alpha$-RuCl$_3$ ($\sim 5.7$ \AA) ~\cite{Ni:arXiv}, suggesting a weaker interlayer coupling in $\alpha$-RuI$_3$ than in $\alpha$-RuCl$_3$. Then, all current experimental information suggest
that $\alpha$-RuI$_3$ is different from $\alpha$-RuCl$_3$.

To better understand the different physical behavior between $\alpha$-RuI$_3$ and $\alpha$-RuCl$_3$, here using the DFT we provide a comprehensive first-principles study of these bulk systems. First, we found the ABC stacking honeycomb-layer $R\overline{3}$ (No. 148) structure is the most likely stacking order of $\alpha$-RuI$_3$ along the $c$-axis. Furthermore, both $R\overline{3}$ and $P\overline{3}1c$ are dynamically stable because no imaginary frequency modes were obtained in the phononic dispersion spectrum. In addition, the $p-d$ hybridization increases from Cl to I, leading to the ''effective'' decrease of the atomic Coulomb repulsion $U$, resulting in the electrons of RuI$_3$ to be less localized than in RuCl$_3$. The effective electronic correlation is reduced in I to a value not large enough to open the spin-orbit Mott gap in $\alpha$-RuI$_3$, leading to its metallic nature. Furthermore, we observed that introducing a large (unrealistic) $U_{\rm eff} = 6$ eV, the spin-orbit Mott gap does open in $\alpha$-RuI$_3$, supporting the consistency of the physical picture we proposed.

\section{II. Calculation Method}

In the present study, we performed first-principles DFT calculations using the projector augmented wave (PAW) method, as implemented in the Vienna {\it ab initio} simulation package (VASP) code~\cite{Kresse:Prb,Kresse:Prb96,Blochl:Prb}. For the electronic correlations, the generalized gradient approximation (GGA) with the Perdew-Burke-Ernzerhof (PBE) potential was employed~\cite{Perdew:Prl} in our DFT calculations. Our plane-wave cutoff energy was $400$ eV. Furthermore, the $k$-point mesh was appropriately modified for different structures to render the in-plane $k$-point densities approximately the same in reciprocal space (e.g., $8\times8\times3$ for the R$\overline{3}$ phase of $\alpha$-RuI$_3$). Note that those $k$-point meshes were tested to confirm that converged energies were produced. Both the lattice constants and atomic positions were fully relaxed until the Hellman-Feynman force on each atom was smaller than $0.01$ eV/{\AA}. The phonon spectra were calculated using the finite displacement approach and analyzed by the PHONONPY software~\cite{Chaput:prb,Togo:sm}. Moreover, on-site Coulomb interactions were considered by using the Dudarev's rotationally invariant DFT+$U$ formulation~\cite{Dudarev:prb} with $U_{\rm eff}= U-J = 2$~eV, where this effective $U_{\rm eff}$ is believed to provide an excellent description of $\alpha$-RuCl$_3$~\cite{Kim:prb16}. It should be noted that hybrid-exchange correlation functionals, such as B3LYP, allow to achieve an excellent agreement with experiments for the band gaps of complex oxide materials~\cite{Eglitis:cry,Eglitis:sys}, whereas the DFT technique usually underestimates the band gaps. The hybrid-exchange correlation functional provides only a correction for the band gap, and does not change other physical properties. However, the scope of this publication is to focus on the physical properties of the metallic phase of RuI$_3$. Hence, our DFT+$U$ calculations are good enough to qualitatively describe the system we focus on. All the crystal structures were visualized with the VESTA code~\cite{Momma:vesta}.

Based on the R$\overline{3}$ (No. 148) structure of $\alpha$-RuI$_3$, we compared the results of optimized crystal structures using different exchange-correlation functionals, with or without vdW interactions, including PBE~\cite{Perdew:Prl}, PBEsol~\cite{Perdew:Prl08}, zero damping vdW-D3~\cite{Grimme:jcp}, and vdW-D3 with Becke-Jonson damping~\cite{Grimme:jcc}.
As shown in Table~\ref{Table1}, all the obtained in-plane lattice constants of different exchange-correlation functionals are close to the experimental values, with the largest discrepancy being $2.4 \%$ for the $a$ value in PBE. But the PBE+D3 with zero damping functional provides the most accurate description for the $c$-axis ($1 \%$ difference from experimental value). Furthermore, the obtained in-plane lattices of PBE+D3 with zero damping functional are only $1.2 \%$ larger than experimental results. Hence, we used the PBE+D3 with zero damping method in the structural optimization of the bulk properties in the rest of the publication.

\begin{table}
\centering\caption{The optimized lattice constants ({\AA}) of the R$\overline{3}$ (No. 148) structure of $\alpha$-RuI$_3$, using the PBE, PBE+D3, PBE+D3, PBEsol, PBEsol+D3, and PBEsol+D3(BJ) methods. The experimental values (Exp. for short) are also listed for comparison, which were reported to form the R$\overline{3}$ (No. 148) structure~\cite{Ni:arXiv}. Note that D3 denotes vdW-D3 with zero damping and D3(BJ) denotes vdW-D3 with  Becke-Jonson damping.}
\begin{tabular*}{0.48\textwidth}{@{\extracolsep{\fill}}lllc}
\hline
\hline & $a$& $b$ & $c$ \\
\hline
PBE      & 6.957  & 6.957 & 20.369    \\
PBE+D3      & 6.875  & 6.875    & 18.841   \\
PBE+D3(BJ)      & 6.817 & 6.817 & 18.301   \\
PBEsol       &  6.821  &  6.821 & 18.487 \\
PBEsol+D3        &  6.761  &  6.761 & 17.918     \\
PBEsol+D3(BJ)     & 6.695 & 6.695 & 17.393     \\
Exp.     & 6.791 & 6.791 & 19.026 \\
\hline
\hline
\end{tabular*}
\label{Table1}
\end{table}

\section{III. Results}

\subsection{A. Stacking order of $\alpha$-RuI$_3$ along the $c$-axis.}
\begin{figure*}
\centering
\includegraphics[width=0.96\textwidth]{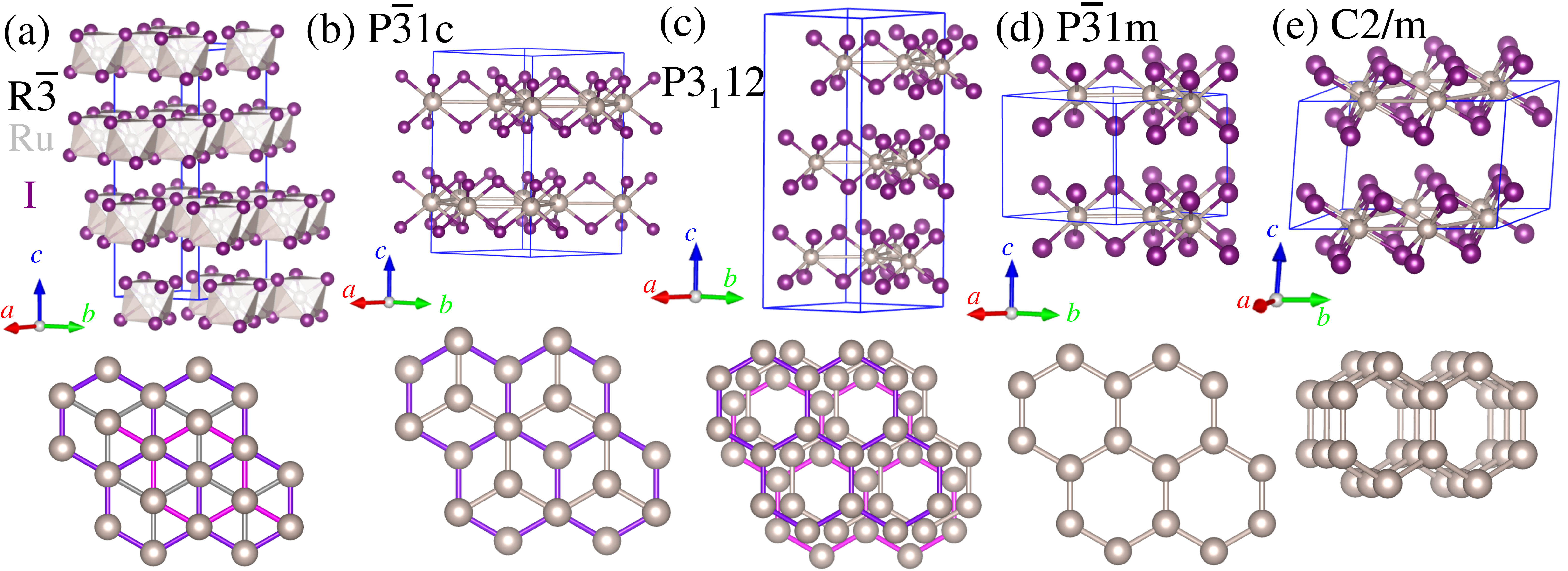}
\caption{Five different conventional cells (gray = Ru; purple = I) with one-, two-, and three-layer periodicity along the $c$-axis. The lower panels display the schematic view of three Ru-honeycomb lattices for crystal structures, respectively. Solid blue lines in the top panels depict conventional cells. (a) $R\overline{3}$ (No. 148), (b) $P\overline{3}1c$ (No. 163), (c) $P3_{1}12$ (No. 151), (d) $P\overline{3}1m$ (No. 162), and (a) $C2/m$ (No. 12).}
\label{Fig1}
\end{figure*}

First, let us discuss the stacking order of $\alpha$-RuI$_3$. Five stacking configurations were considered in our study, shown in Fig.~\ref{Fig1}, where the lower panels display the top view of the Ru-honeycomb sublattice. The main difference between those five structures is the different stacking ordering along the $c$-axis. The $R\overline{3}$, $P3_{1}12$, and $C2/m$ structures involve three-layer periodicity stacking, resulting in three Ru-honeycomb sublattices stacking, as shown in the lower panels in Fig.~\ref{Fig1}. Note that the monoclinic $C2/m$ structure was reported to be the crystal structure of $\alpha$-RuCl$_3$~\cite{Johnson:prb15} with in-plane shifting of the Ru-honeycomb stacking, and it is similar to the space group of $P3_{1}12$ that was also suggested as the space group of $\alpha$-RuCl$_3$~\cite{Kubota:prb15}. The $P\overline{3}1m$ and $P\overline{3}1c$ structures involve only one- and two-layer periodicity along the $c$-axis. Note that both $R\overline{3}$ and $P\overline{3}1c$ structures were shown experimentally to be the space group of $\alpha$-RuI$_3$~\cite{Ni:arXiv,Nawa:arXiv}. Furthermore, the Ru-Ru dimerization was reported experimentally in the $C2/m$ and $P3_{1}12$ structures of $\alpha$-RuCl$_3$~\cite{Johnson:prb15,Kubota:prb15}.

Our optimized lattice constants are $a = b = 6.875$, $c = 18.841$~\AA~and $a = b = 6.873$, $c = 12.577$~\AA~for $R\overline{3}$ (No. 148) and $P\overline{3}1c$ (No. 163), respectively, which are close to the experimental results ($a = b = 6.791$, $c = 19.026$~\AA~for $R\overline{3}$ and $a = b = 6.778$, $c = 12.579$~\AA~for $P\overline{3}1c$)~\cite{Ni:arXiv,Nawa:arXiv}. Based on the optimized structures, we calculated their relative total energies with GGA in the nonmagnetic state. We found that the $R\overline{3}$ (No. 148) configuration has the lowest energy, indicating this stacking structure is the most likely stacking order among all the candidates. Note that here we do not consider nuclear quantum effects. Based on our results, then the proper conclusion is that $R\overline{3}$ (No. 148) is the most possible ground state in the {\it absence} of nuclear quantum effects. However, this could change with the inclusion of those nuclear quantum effects, and consequently a final determination is left to future work. The $P\overline{3}1c$ structure has a slightly higher energy than the $R\overline{3}$ structure.

\begin{table}
\centering\caption{Optimized lattice constants ({\AA}) and energy differences (meV/Ru) with respect to the  $R\overline{3}$ (No. 148) configuration taken as the
reference of energy, for the various structural configurations. The experimental values (Exp. for short) are also listed for comparison, which were reported to form the $R\overline{3}$ (No. 148)~\cite{Ni:arXiv} and $P\overline{3}1c$ (No. 163) structures~\cite{Nawa:arXiv}, respectively. }
\begin{tabular*}{0.48\textwidth}{@{\extracolsep{\fill}}llllc}
\hline
\hline
  & a& b & c  & Energy \\
\hline
$R\overline{3}$ (No. 148)      & 6.875  & 6.875    & 18.841  & 0   \\
$P\overline{3}1c$ (No. 163)       & 6.873 & 6.873 & 12.577 & 2.58 \\
$P3_{1}12$ (No. 151)      & 6.841 & 6.841 & 19.209  & 6.35 \\
$P\overline{3}1m$ (No. 162)      & 6.828 & 6.828 & 6.413 & 25.83 \\
$C2/m$ (No. 12)       & 6.865 & 11.827 & 6.764 & 7.68    \\
Exp.~\cite{Ni:arXiv}     & 6.791 & 6.791 & 19.026 & --   \\
Exp.~\cite{Nawa:arXiv}     & 6.778 &6.778  & 12.579 &--\\
\hline
\hline
\end{tabular*}
\label{Table2}
\end{table}

\begin{figure}
\centering
\includegraphics[width=0.48\textwidth]{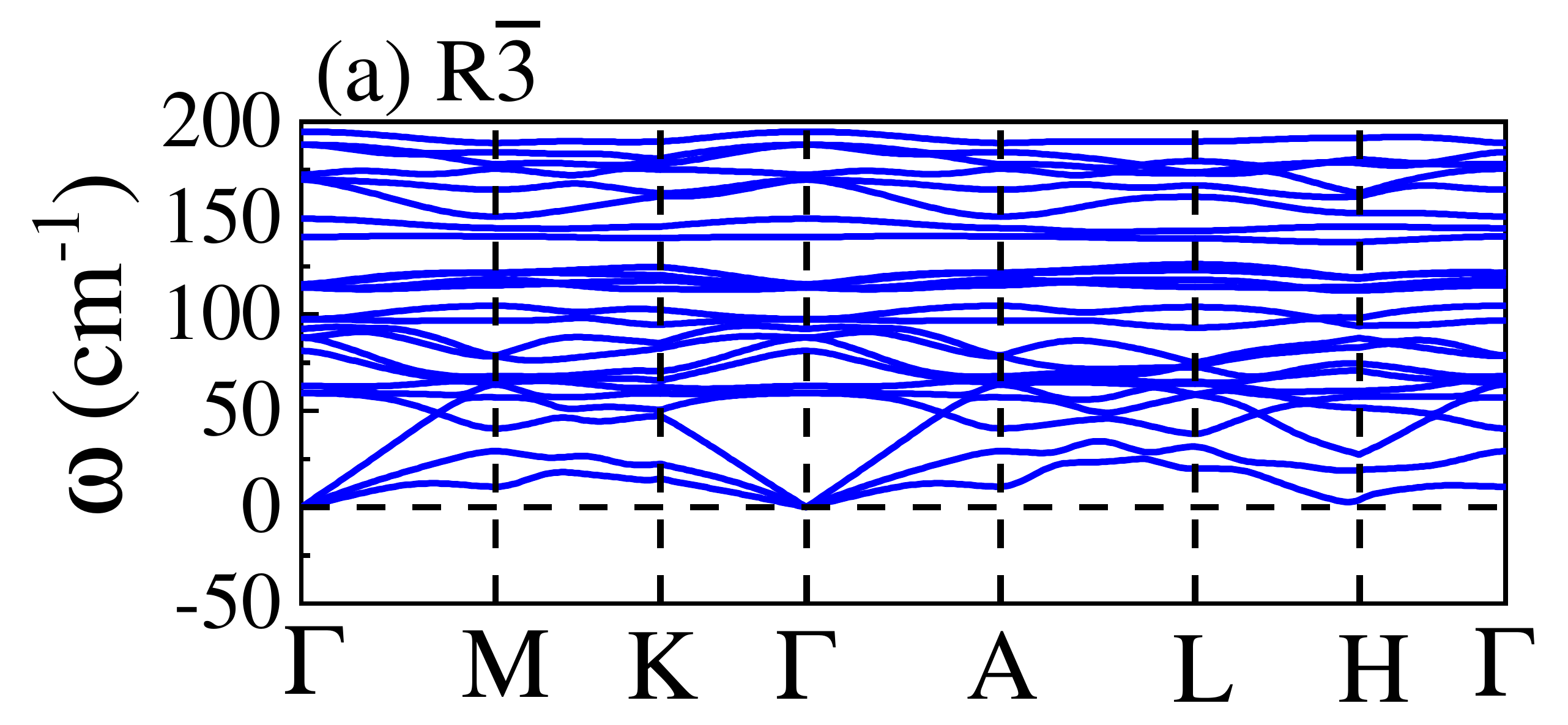}
\includegraphics[width=0.48\textwidth]{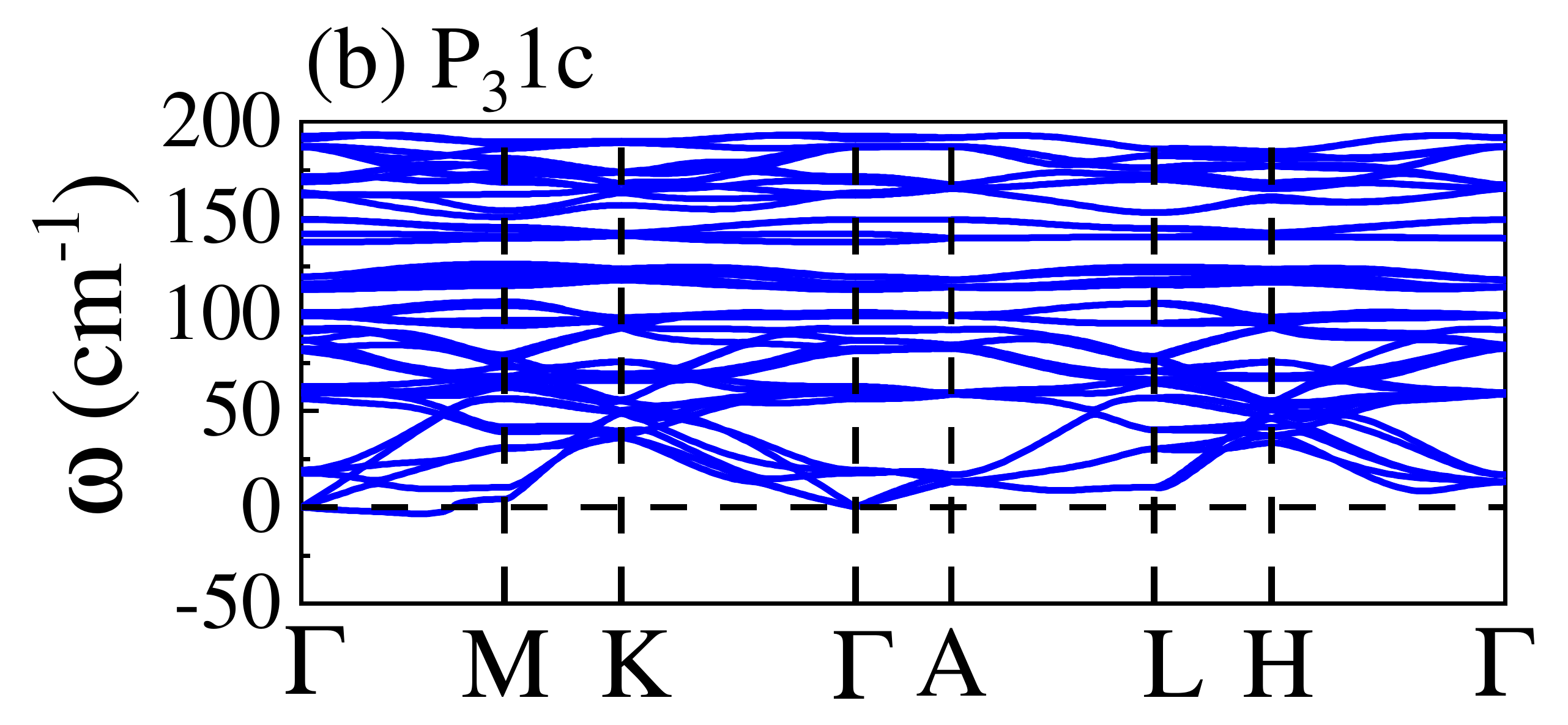}
\caption{The calculated phonon spectrum of $\alpha$-RuI$_3$ for the structures (a) $R\overline{3}$ and (b) $P\overline{3}1c$, respectively, in the nonmagnetic state without Hubbard $U$. The coordinates of the high-symmetry points in the bulk Brillouin zone (BZ) are $\Gamma$ = (0, 0, 0), M = (0.5, 0, 0), K = (1/3, 1/3, 0), A = (0, 0, 0.5), L = (0.5, 0, 0), and H = (1/3, 1/3, 0.5).}
\label{Fig2}
\end{figure}

To better understand the structural stability of $\alpha$-RuI$_3$, we carried out the phononic dispersion calculations using a $2\times2\times1$ supercell for the $R\overline{3}$ (No. 148) and $P\overline{3}1c$ (No. 163) phases, respectively. Figure~\ref{Fig2} indicates that the $R\overline{3}$ and $P\overline{3}1c$ structures are dynamically stable because no imaginary frequency modes were obtained in the phononic dispersion spectrum.

In addition, we also considered the Coulombic repulsion $U$ effect on the process of optimizing crystal lattices (see Table.S1). The lattice structures do not change much compared to the lattice structures without $U_{\rm eff}$ and the $R\overline{3}$ structure of $\alpha$-RuI$_3$ has the lowest energy among those five lattice configurations. Furthermore, we also compared the energies between the $R\overline{3}$ and $P\overline{3}1c$ structures with and without the effective Coulomb repulsion $U$ effect, considering the experimentally reported lattice structures where the $R\overline{3}$ $\alpha$-RuI$_3$ always has lower energy than $P\overline{3}1c$ $\alpha$-RuI$_3$. Moreover, we also calculated the phononic dispersion with $U_{\rm eff} = 2.0$ eV for both the $R\overline{3}$ and $P\overline{3}1c$ phases. We found that the $R\overline{3}$ phase is dynamically stable because no imaginary frequency modes were obtained in the phononic dispersion spectrum. However, the phononic dispersion spectrum of $P\overline{3}1c$ suggests that this structure is unstable, as displayed in Fig. ~S1b. In this case, the electronic correlation effects may induce an structural phase transition for the $P\overline{3}1c$ case. Since the energy difference of these two structures is quite small, a possible structural phase transition at finite temperatures deserves further experimental investigation and discussion beyond the scope of our present manuscript. Hence, based on our DFT calculations, we believe that the $R\overline{3}$ structure is the most likely crystal structure of $\alpha$-RuI$_3$. It should also be noted that $R\overline{3}$ and $P\overline{3}1c$ have quite similar crystal and electronic structures.
The metallic behavior and strong $p-d$ hybridizations are also obtained in the $P\overline{3}1c$ structure of $\alpha$-RuI$_3$ (see Figs.~S3-S4).

For the benefit of the readers, we also present here the corresponding electronic structures of $P\overline{3}1c$ $\alpha$-RuI$_3$ in the Supplementary Material (SM)~\cite{Supplemental}. We also remark that the main physical conclusion of our manuscript is not affected by the structural configurations because the difference between those structures is the stacking arrangement along the $c$-axis. In the rest of the text, we will focus on discussing the results of $R\overline{3}$ $\alpha$-RuI$_3$ starting in the next section.

\subsection{B. Electronic structures.}

\begin{figure}
\centering
\includegraphics[width=0.48\textwidth]{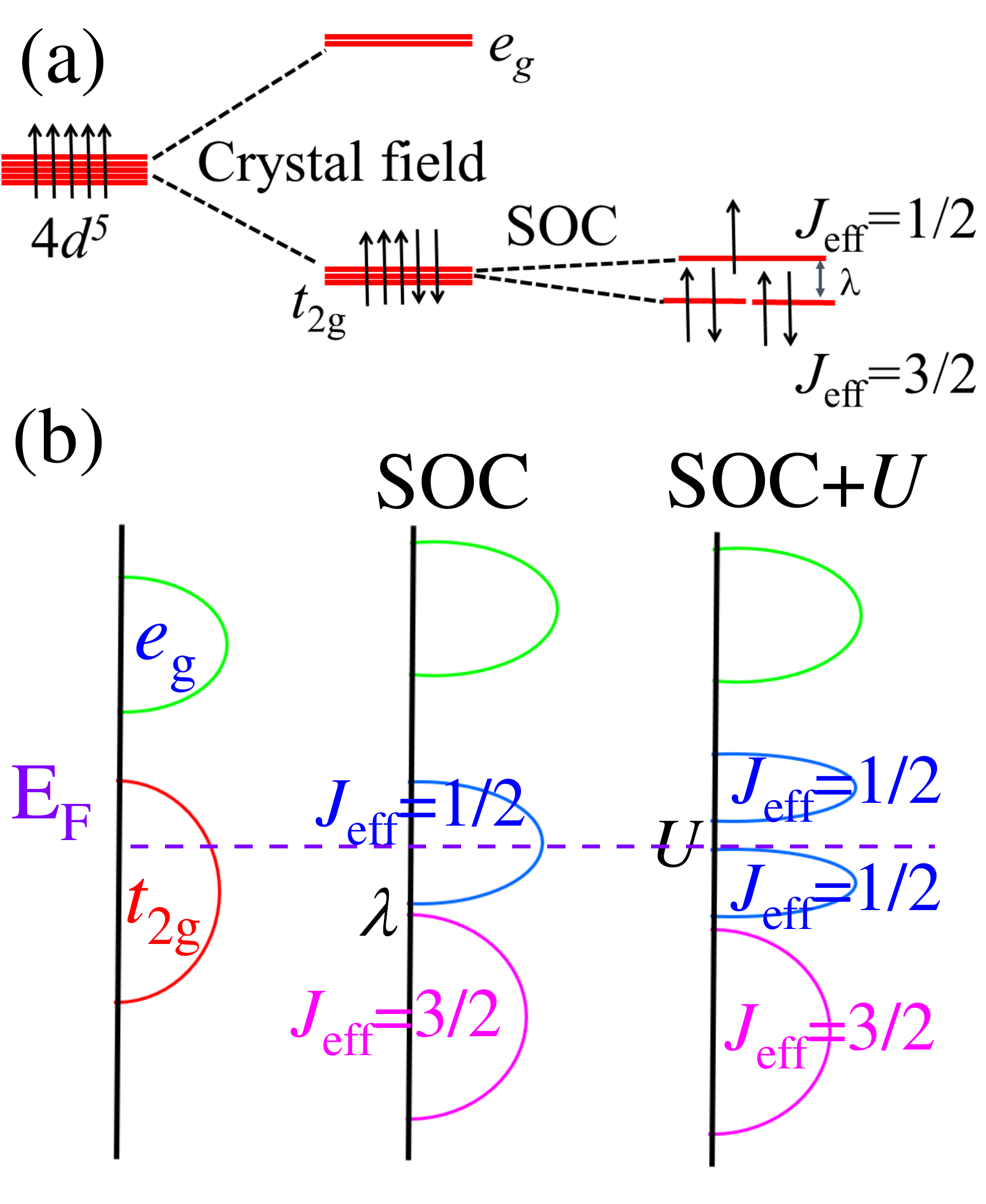}
\caption{ (a) Schematic energy splitting of the Ru $4d$ orbitals in the $d^5$ low-spin configuration. (b) Schematic of the local density-of-states for the cases without interactions, with SOC only, and with both SOC+$U$, in this Ru $4d^5$ configuration.}
\label{Fig3}
\end{figure}

Let us now discuss the energy splitting of the Ru's $4d^5$ orbitals, as sketched in Fig.~\ref{Fig3}(a). First, the crystal field leads to three lower-degenerate-energy $t_{2g}$ orbitals ($d_{xy}$, $d_{yz}$, and $d_{xz}$) and two higher-degenerate-energy $e_g$ orbitals ($d_{x^2-y^2}$ and $d_{3z^2-r^2}$). In addition, by introducing the SOC effect, the three lower-degenerate-energy $t_{2g}$ orbitals split into two energy states $J_{\rm eff} = 3/2$ and $J_{\rm eff} = 1/2$. The Ru$^{\rm 3+}$ state is considered a $d^5$ electronic configuration with a low-spin state. Thus, this system could be regarded as a $J = 1/2$ (half-occupied $J_{\rm eff} = 1/2$ state), while the two $J_{\rm eff} = 3/2$ states are fully occupied, as shown in Fig.~\ref{Fig3}(a).

In general, the density of states (DOS) of this $4d^5$ low-spin configuration could be intuitively understood as displayed in Fig.~\ref{Fig3}(b). Under a cubic crystal field, the five $4d$ electrons of Ru populate the lower $t_{\rm 2g}$ bands separated by the crystal-field splitting energy ($~\sim 10$ Dq), resulting in a metallic phase because the $t_{\rm 2g}$ orbitals are not completely occupied. Then, by introducing the SOC
effect, the $J_{\rm eff} = 1/2$ and $J_{\rm eff} = 3/2$ states begin to separate from each other, leading to a half-occupied $J_{\rm eff} = 1/2$ state and two fully-occupied $J_{\rm eff} = 3/2$ states, where the splitting energy depends on the SOC strength $\lambda$. In this case, the system is still metallic since the $J_{\rm eff} = 1/2$ is not completely occupied. Finally, increasing the on-site electronic correlations $U$ leads to an energy gap for the $J_{\rm eff} = 1/2$ band near its Fermi surface as well, resulting in a Mott transition. In this case, this insulating gap system is also often referred to as the ``spin-orbit Mott insulating'' gap.

\begin{figure}
\centering
\includegraphics[width=0.48\textwidth]{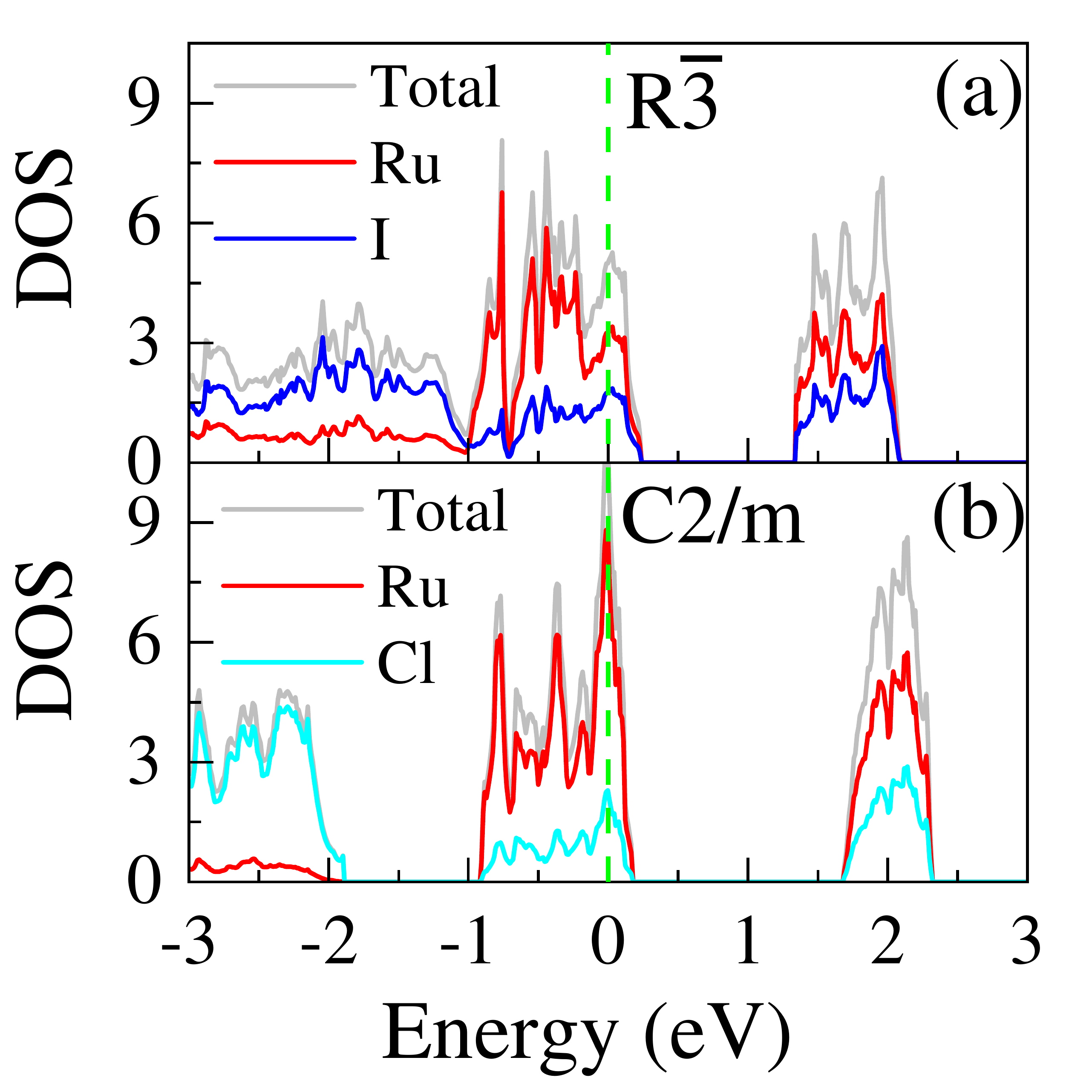}
\includegraphics[width=0.48\textwidth]{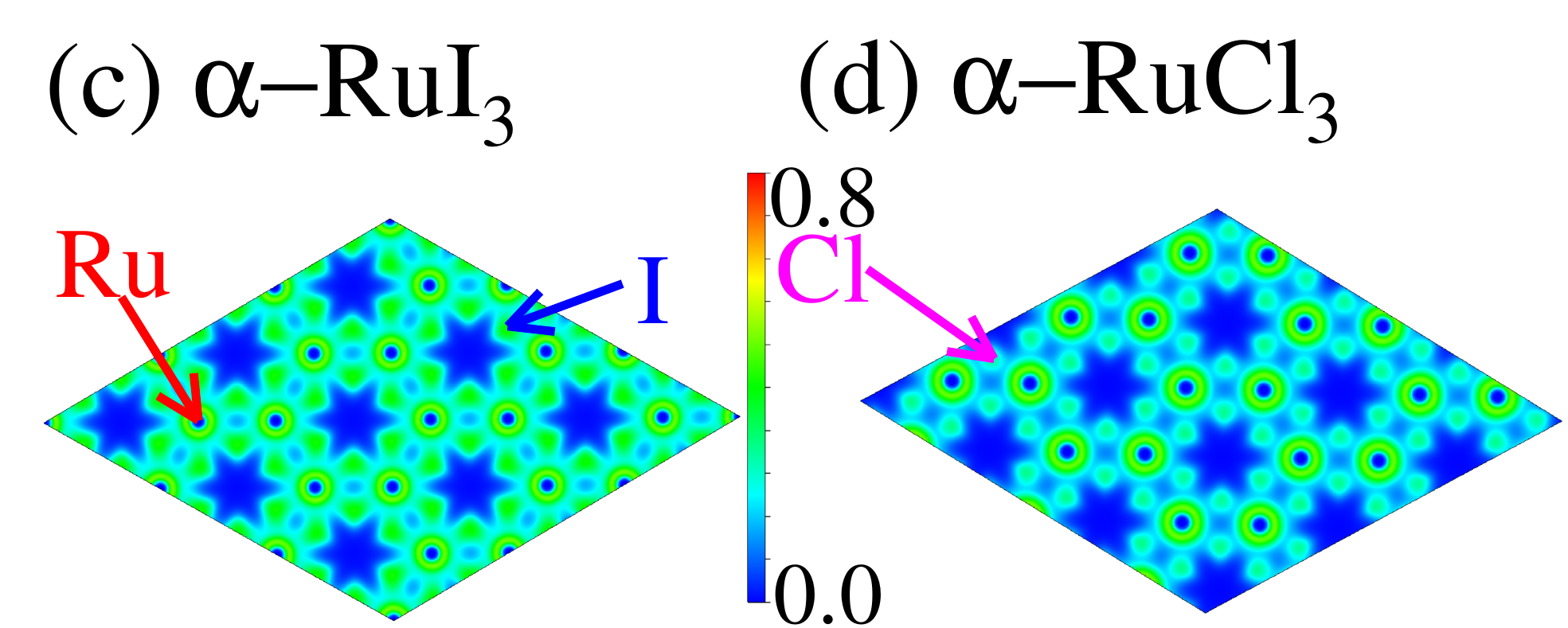}
\caption{(a-b) Density of states near the Fermi level based on the nonmagnetic states for $\alpha$-RuI$_3$ and $\alpha$-RuCl$_3$, respectively. Gray: Total; red: Ru; blue: I; cyan: Cl. The Fermi level is marked by the vertical dashed green line. (a) Results for the $R\overline{3}$ structure of $\alpha$-RuI$_3$. (b) Results for the $C2/m$ structure of $\alpha$-RuCl$_3$. (c-d) Electron localization function of one Ru-honeycomb layer for $\alpha$-RuI$_3$ and $\alpha$-RuCl$_3$, respectively, corresponding to nonmagnetic phases, in the $a-b$ plane.}
\label{Fig4}
\end{figure}

To better understand the similarities and differences between $\alpha$-RuI$_3$ and $\alpha$-RuCl$_3$, we calculated the DOS of $\alpha$-RuI$_3$ with the $R\overline{3}$ structure and $\alpha$-RuCl$_3$ with the $C2/m$ structure, both for the nonmagnetic phase, respectively. According to the calculated DOS [see Figs.~\ref{Fig4}(a-b)], the bands near the Fermi level are mainly contributed by the Ru-$4d$ $t_{\rm 2g}$ orbitals, $hybridized$ with the I-$5p$ and Cl-$3p$ orbitals, respectively. Furthermore, the I-$5p$ orbitals are closer to the Fermi level than the Cl-$3p$ orbitals, as shown in Figs.~\ref{Fig4}(a-b). With increasing atomic radius from Cl to I, the $p$ components near the Fermi level become larger, leading to an increase in the $p-d$ hybridization tendency from I to Cl. In addition, the low-energy $t_{2g}$ bands are more extended in $\alpha$-RuI$_3$ than in $\alpha$-RuCl$_3$, indicating stronger electronic correlations ($U/W$, where $W$ is the bandwidth) in the $\alpha$-RuCl$_3$ case. To open the Mott gap in the $J_{\rm eff} = 1/2$ state, $\alpha$-RuI$_3$ needs a larger Coulomb repulsion $U$ than $\alpha$-RuCl$_3$. It should be noted that those results are obtained in the $P\overline{3}1c$ structure of $\alpha$-RuI$_3$ and the $P3_{1}12$ structure of $\alpha$-RuCl$_3$ (see SM)~\cite{Supplemental}.

In addition, we also calculated the electron localization function (ELF)~\cite{Savin:Angewandte} for the the $\alpha$-RuI$_3$ and $\alpha$-RuCl$_3$ cases, respectively, as displayed in Figs.~\ref{Fig4}(c-d).
The ELF picture indicates that the charges are less localized inside the Ru-I bonds, resulting in large hybridized $p-d$ bonds in $\alpha$-RuI$_3$, in contrast to the localized charges along with the Ru-Cl bonds in $\alpha$-RuCl$_3$. The movement of electrons is by tunneling from Ru to I (or Cl), and then to another Ru. In other words, iodine (or chlorine) is the bridge between rutheniums. Hence, it is easy to imagine that RuCl$_3$ is more Mott-localized than RuI$_3$ by using the same value of the on-site repulsion $U$ at the Ru site. The reason is that comparing against RuCl$_3$, the bandwidth of RuI$_3$ is increased, indicating that the electronic correlation $U/W$ has decreased. In this case, due to the increase in the $p-d$ hybridization of $\alpha$-RuI$_3$, the ``effective'' Coulomb repulsion $U/W$ will decrease in $\alpha$-RuI$_3$, reducing or directly not even opening an energy gap. Hence, RuI$_3$ displays metallic behavior, in contrast to the insulating behavior in $\alpha$-RuCl$_3$.

\begin{figure}
\centering
\includegraphics[width=0.48\textwidth]{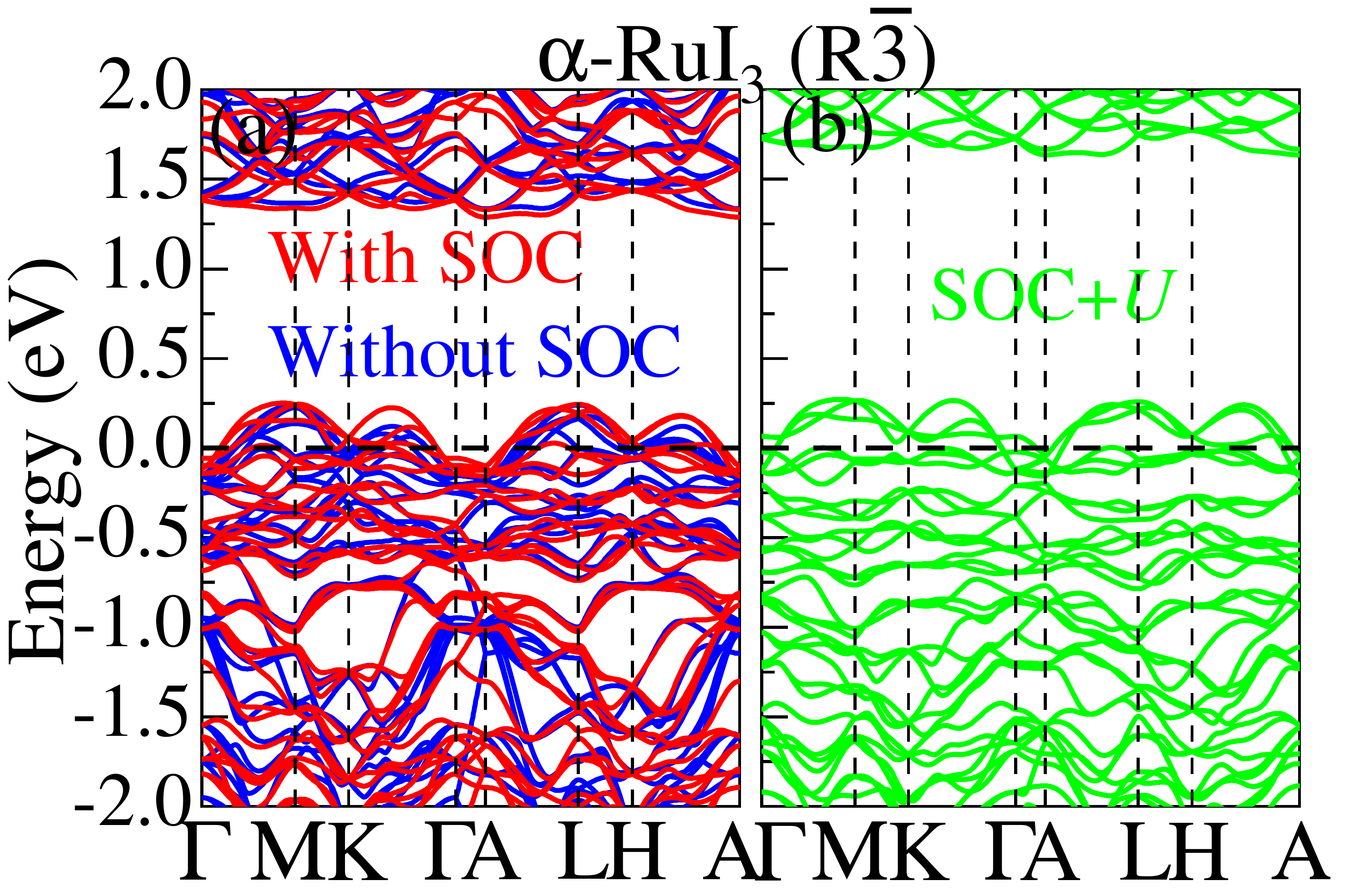}
\caption{Calculated electronic bands structures of $\alpha$-RuI$_3$ in the $R\overline{3}$ structure using a nonmagnetic state: (a) without/with SOC (color convention indicated) and (b) with SOC plus $U_{\rm eff} = 2$ eV, respectively. The coordinates of the high-symmetry points in the bulk Brillouin zone (BZ) are $\Gamma$ = (0, 0, 0), M = (0.5, 0, 0), K = (1/3, 1/3, 0), A = (0, 0, 0.5), L = (0.5, 0, 0), and H = (1/3, 1/3, 0.5).}
\label{Fig5}
\end{figure}

\begin{figure}
\centering
\includegraphics[width=0.48\textwidth]{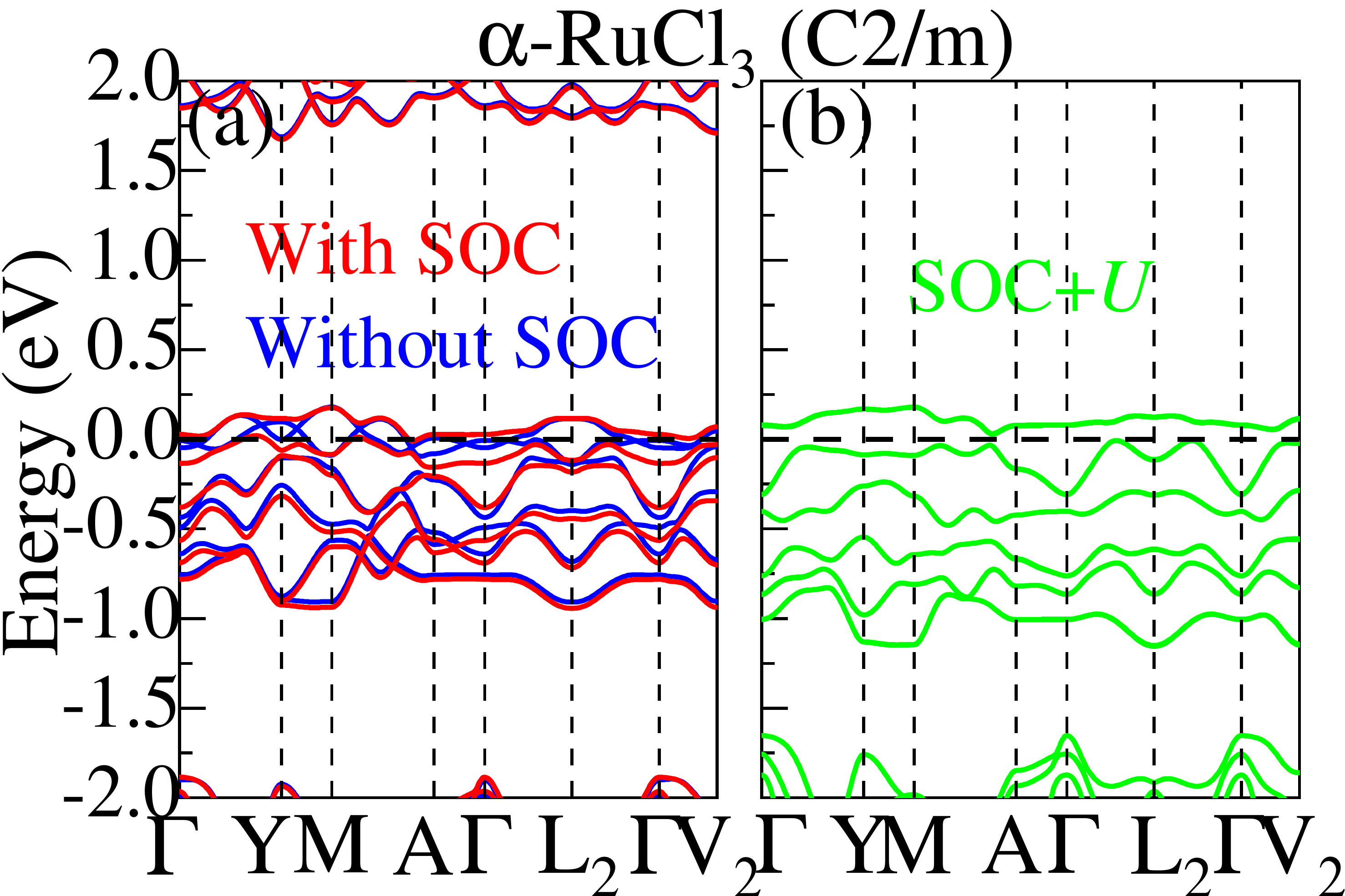}
\caption{Calculated electronic band structures of $\alpha$-RuCl$_3$ with the $C2/m$ structure in the nonmagnetic
state: (a) without/with SOC (color convention indicated) and (b) with SOC plus $U_{\rm eff} = 2$ eV, respectively. The coordinates of the high-symmetry points in the bulk BZ are $\Gamma$ = (0, 0, 0), Y = (0.5, 0.5, 0), M = (0.5, 0.5, 0.5), A = (0, 0, 0.5), L$_2$ = (0, 0.5, 0.5), and V$_2$ = (0, 0.5, 0).}
\label{Fig6}
\end{figure}

Furthermore, we calculated band structures of $R\overline{3}$ $\alpha$-RuI$_3$ with and without SOC effect and with the effective Coulomb repulsion $U$ ($U_{\rm eff} = 2$ eV). As shown in Fig.~\ref{Fig5}, both non-SOC and SOC-calculated band structures suggest metallic behavior in $\alpha$-RuI$_3$. Taking into account SOC and the Coulomb repulsion $U$ ($U_{\rm eff} = 2$ eV), $\alpha$-RuI$_3$ {\it still} displays metallic behavior of the $J_{\rm eff} = 1/2$ bands but opens gaps on some high symmetry points in the Brillouin zone, as displayed in Fig.~\ref{Fig5}(b). Those results are consistent with our previous analysis that the ``effective'' Coulomb repulsion would reduce or directly not open the gap in $\alpha$-RuI$_3$. For comparison, we also calculated the band structure of
$\alpha$-RuCl$_3$ using the $C2/m$ structure with and without SOC effect and with the Coulomb repulsion $U$ ($U_{\rm eff} = 2$ eV). As displayed in Fig.~\ref{Fig6}, the band structure clearly shows insulating behavior for the $J_{\rm eff} = 1/2$ bands with a Mott transition caused by the Coulomb repulsion $U$. Based on our estimation, the spin-orbital couplings are about $0.12$ and $0.2$ eV for RuCl$_3$ and RuI$_3$, respectively, in agreement with other theoretical studies~\cite{Kim:prb15,Nawa:arXiv}. Hence, after using suitable parameters, we obtained metallic behavior in $\alpha$-RuI$_3$ and insulating behavior in $\alpha$-RuCl$_3$, in excellent agreement with the experimental results. This can be naturally explained in simple terms: increasing the $p-d$ hybridization of $\alpha$-RuI$_3$ decreases the effective electronic correlations $U/W$ because the bandwidth $W$ increases, and thus allows for the conduction of charge along with the Ru-I bonds.

\subsection{C. Comparing $\alpha$-RuI$_3$ with $\alpha$-RuCl$_3$ using zigzag AFM order}

The preliminary experimental characterization of $\alpha$-RuI$_3$ reveals the absence of long-range magnetic order down to $0.35$ K, suggesting a paramagnetic metallic state~\cite{Ni:arXiv,Nawa:arXiv}. On the contrary, $\alpha$-RuCl$_3$ is in a spin-orbital Mott state with a zigzag AFM ordering in the ground state at low temperatures~\cite{Koitzsch:prb16,Cao:prb16}. To better understand the different conductive behavior of $\alpha$-RuI$_3$ and $\alpha$-RuCl$_3$, we calculated the electronic structures for the two materials in both assuming zigzag AFM order. Because here we are simply performing a qualitative analysis of the effect of the Coulomb repulsion $U$, we used the $C2/m$ symmetry for the crystal structure for both $\alpha$-RuI$_3$ and $\alpha$-RuCl$_3$.

\begin{figure}
\centering
\includegraphics[width=0.48\textwidth]{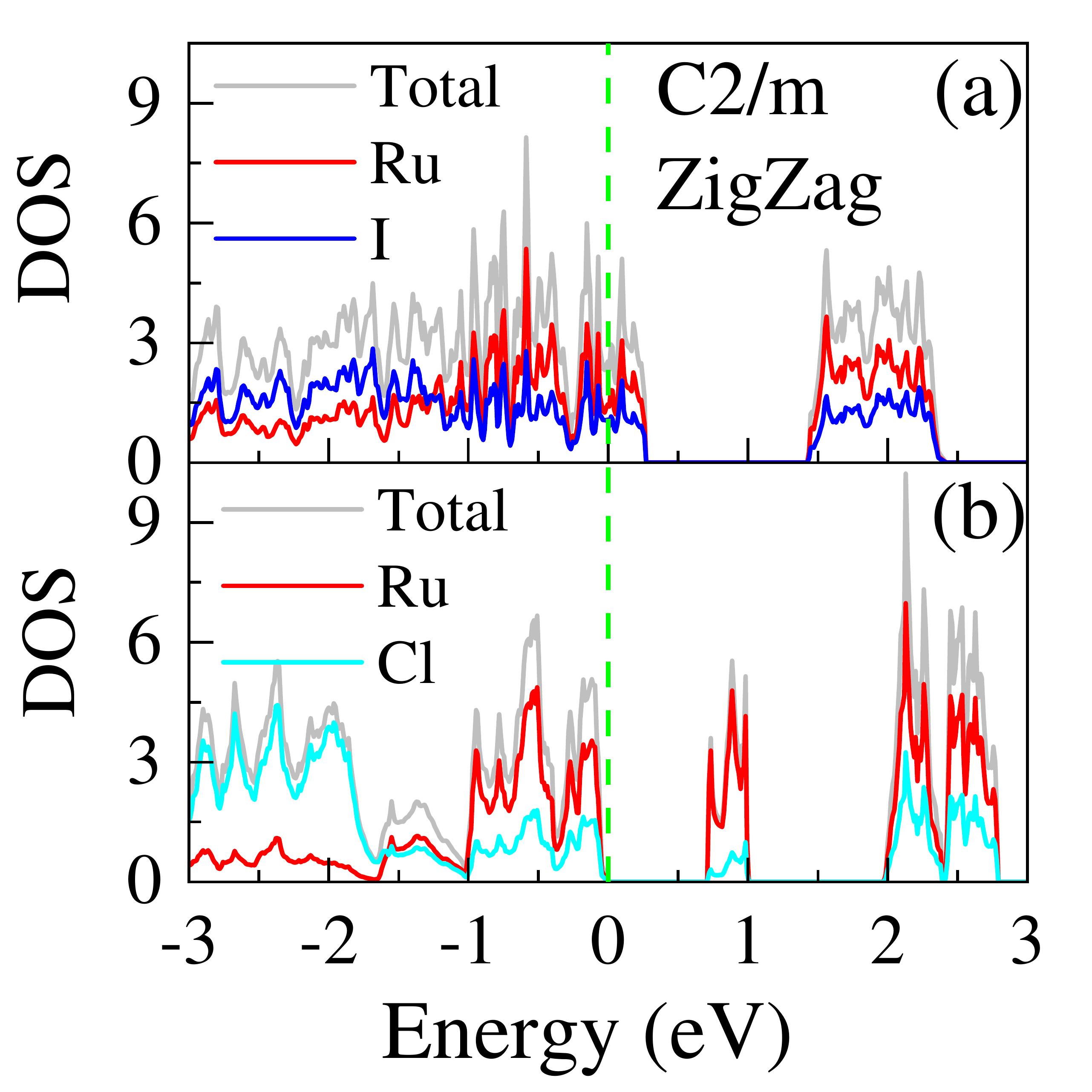}
\caption{Density of states near the Fermi level for the zigzag AFM state using the $C2/m$ structure for both $\alpha$-RuI$_3$ and $\alpha$-RuCl$_3$, and with SOC plus $U$ ($U_{\rm eff} = 2$ eV) also for both materials as well. Gray: Total; red: Ru; blue: I; cyan: Cl. The Fermi level is marked by the green dashed line.  (a) $\alpha$-RuI$_3$, where the Fermi level is inside the valence band and (b) $\alpha$-RuCl$_3$ where the Fermi level is inside the gap.}
\label{Fig7}
\end{figure}

Based on previous studies~\cite{Pollini:prb96,Plumb:prb14,Kim:prb15,Koitzsch:prb16}, the $U_{\rm eff}$ has been estimated to be about $1 - 2$ eV for Ru atoms, which are often used in the band structure calculations of trihalogen ruthenium compounds. In addition, the effective $U_{\rm eff} = 2.0$ eV is believed to provide an excellent description of the stacking order of RuCl$_3$~\cite{Kim:prb16}. Hence, we used $U_{\rm eff} = 2.0$ eV in our magnetic calculations. Note that we also tested other values of $U_{\rm eff}$, but they do not change our main conclusion~\cite{Supplemental}. Furthermore, for RuCl$_3$, previous optical data found a small optical gap about $0.3$ eV~\cite{Binotto:pss}, but this very small value was considered not to be associated with charge excitations~\cite{Sandilands:prb16}. Moreover, some other experiments suggested that the optical gap was around $1$ eV~\cite{Plumb:prb14,Sandilands:prb16}. Hence, our results are in good agreement with the optical data of $\alpha$-RuCl$_3$ qualitatively. Figure~\ref{Fig7} indicates that $\alpha$-RuI$_3$ still displays metallic behavior, in contrast to the insulating behavior in $\alpha$-RuCl$_3$. The same SOC+$U$ ($U_{\rm eff} = 2$ eV) opens a gap ($\sim 0.7$ eV) in $\alpha$-RuCl$_3$, but could not open the Mott gap in $\alpha$-RuI$_3$. This result supports the notion that the effective electronic correlation $U/W$ has been reduced from Cl to I and it is not enough to open a gap, leading to metallic behavior in $\alpha$-RuI$_3$. In this case, the results obtained, even involving the effect of Coulomb repulsion, can naturally explain the metallic behavior in $\alpha$-RuI$_3$, in contrast to the spin-orbit Mott insulating behavior observed in $\alpha$-RuCl$_3$.

\begin{figure}
\centering
\includegraphics[width=0.48\textwidth]{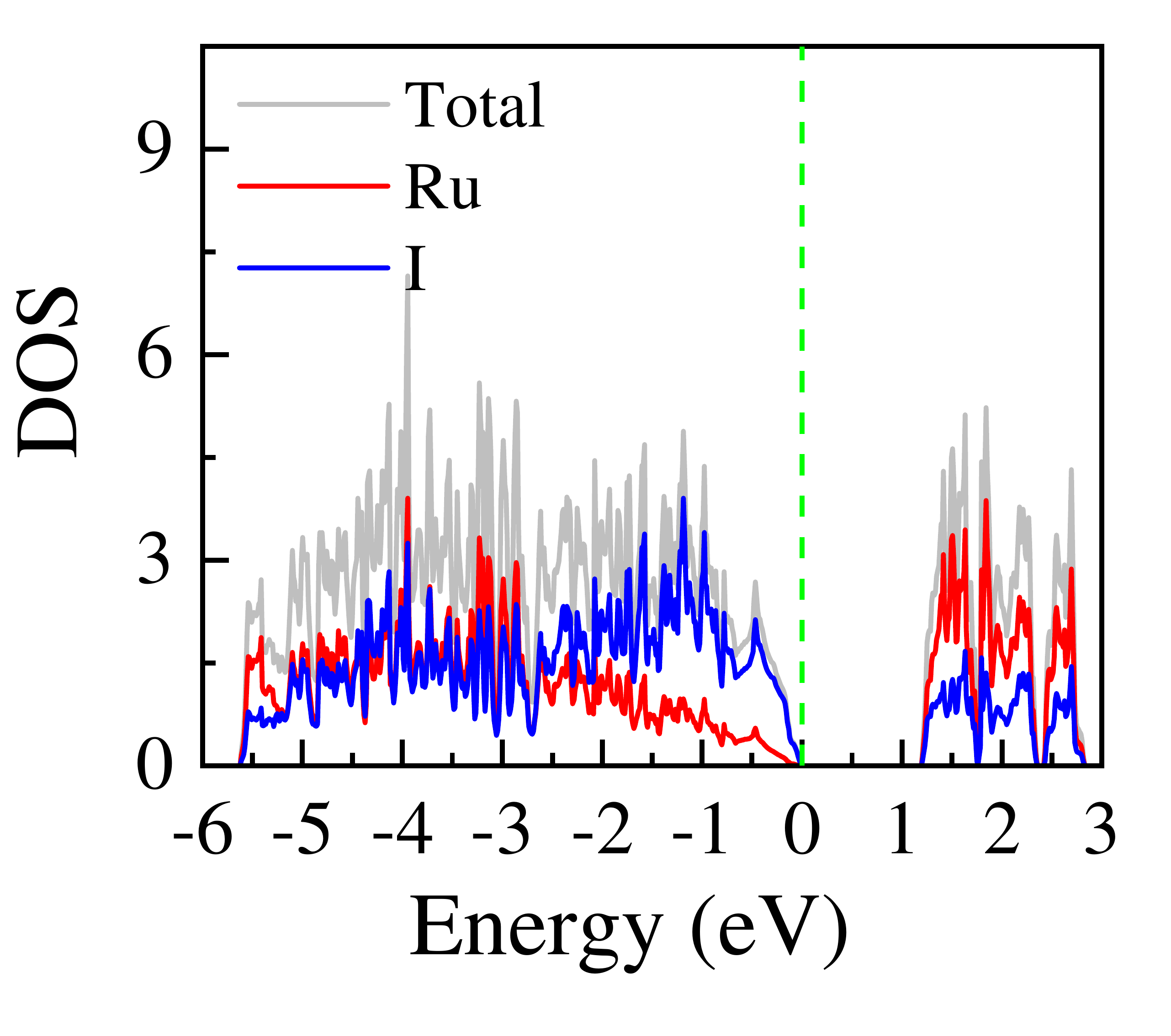}
\caption{Density of states near the Fermi level using a zigzag AFM state and the structure $C2/m$ for $\alpha$-RuI$_3$ with GGA+SOC+$U$ (employing an unphysically large $U_{\rm eff} = 6$ eV). Gray: Total; red: Ru; blue: I. The Fermi level is marked by the green dashed vertical line. }
\label{Fig8}
\end{figure}

As discussed in the previous sections, the metallic nature of $\alpha$-RuI$_3$ is induced by the reduced ``effective'' Coulomb repulsion when moving from Cl to I. In essence, the spin-orbit Mott gap opens if the $U$ is large enough. In this case, the Coulomb repulsion $U$ of the Ru atoms shifts Ru states to lower energies and reduces the $p-d$ hybridization, and thus its bandwidth. To confirm this physical picture, we introduced an artificially large (unrealistic) $U_{\rm eff} = 6$ eV on the Ru sites. This $U_{\rm eff}$ is too large for RuI$_3$. As expected, using the same lattice of Fig.~\ref{Fig7}(a), a large spin-orbit Mott gap ($\sim 1.1$ eV) emerges this time in the DOS, as displayed in Fig.~\ref{Fig8}. These results support our physical picture for the explanation of the metallic behavior in RuI$_3$. Note that here we only did a qualitative analysis for the metallic-insulating transition of RuI$_3$ because finding the specific critical value of the
Hubbard repulsion $U$ is also affected by many other aspects besides $U$, such as the lattice structure, magnetic ordering, spin orientation, etc. However, our results are qualitatively sufficient to show that for a large enough $U_{\rm eff}$
a spin-orbit Mott gap opens even in $\alpha$-RuI$_3$.

\section{IV. Conclusions}

In this publication, we have systematically studied the properties of $\alpha$-RuI$_3$ and $\alpha$-RuCl$_3$ by using first-principles DFT. We found that the most likely stacking order of $\alpha$-RuI$_3$ along the $c$-axis is the ABC stacking honeycomb-layer $R\overline{3}$ (No. 148) structure. In addition, both $R\overline{3}$ and $P\overline{3}1c$ were found to be dynamical stable because no imaginary frequency modes were obtained in the phononic dispersion spectrum. By introducing GGA+SOC+$U$ calculations, the $J_{\rm eff} =1/2$ physics was obtained in both $\alpha$-RuI$_3$ and $\alpha$-RuCl$_3$. Different from the spin-orbit Mott insulating phase of $\alpha$-RuCl$_3$, on the other hand $\alpha$-RuI$_3$ displays a strong metallic behavior, in agreement with the currently available experimental information. The strong hybridization between the Ru $4d$ and I $5p$ orbitals decreases the ``effective'' atomic Coulomb repulsion $U/W$, namely increases the bandwidth $W$ in the ratio $U/W$. This effective electronic correlation $U/W$ is reduced from Cl to I, inducing metallic behavior in $\alpha$-RuI$_3$. In our study, by considering a large (unrealistic) $U_{\rm eff} = 6$ eV, the spin-orbit Mott gap finally opens in $\alpha$-RuI$_3$, supporting the physical picture we proposed. In summary, while the atomic $U$ of Ru must be very similar in both compounds,
the bandwidth $W$ in the case of Cl is smaller than in I, and this is sufficient to place $\alpha$-RuCl$_3$ on the insulating
side of the metal-insulator transition, while $\alpha$-RuI$_3$ is still on the metallic side.

\section{Acknowledgments}
The work of Y.Z., L.-F.L., A.M., and E.D. was supported by the U.S. Department of Energy (DOE), Office of Science, Basic Energy Sciences (BES), Materials Sciences and Engineering Division. All the calculations were carried out at the Advanced Computing Facility (ACF) of the University of Tennessee Knoxville (UTK).


\begin{references}
\bibitem{Bednorz:Cu} J. G. Bednorz and K. A. M{\"u}ller, \href{https://doi.org/10.1007/BF01303701}{Z. Phys. B: Condens. Matter \textbf{64}, 189 (1986).}
\bibitem{Dagotto:rmp94} E. Dagotto, \href{https://doi.org/10.1103/RevModPhys.66.763}{Rev. Mod. Phys. \textbf{66}, 763 (1994).}
\bibitem{Stewart:Rmp} G. R. Stewart, \href{https://doi.org/10.1103/RevModPhys.83.1589}{Rev. Mod. Phys. \textbf{83}, 1589 (2011).}
\bibitem{Dagotto:rp} E. Dagotto, T. Hotta, and A. Moreo, \href{https://doi.org/10.1016/S0370-1573(00)00121-6}{Phys. Rep. \textbf{344}, 1 (2001).}
\bibitem{Dagotto:science05} E. Dagotto, \href{https://doi.org/10.1126/science.1107559}{Science {\bf 309}, 257 (2005)}.
\bibitem{Grioni:JPCM} M. Grioni, S. Pons and E. Frantzeskakis, \href{https://doi.org/10.1088/0953-8984/21/2/023201}{J. Phys.: Condens. Matter \textbf{21}, 023201 (2009).}
\bibitem{Scalapino:Rmp} D. J. Scalapino \href{https://doi.org/10.1103/RevModPhys.84.1383}{Rev. Mod. Phys. \textbf{84}, 1383 (2012).}
\bibitem{Dagotto:Rmp} E. Dagotto, \href{https://doi.org/10.1103/RevModPhys.85.849}{Rev. Mod. Phys. \textbf{85}, 849 (2013).}
\bibitem{Monceau:ap} P. Monceau, \href{https://doi.org/10.1080/00018732.2012.719674}{Adv. Phys. \textbf{61}, 325 (2012).}
\bibitem{Lin:prm19} L. F. Lin, Y. Zhang, A. Moreo, E. Dagotto, and S. Dong, \href{https://doi.org/10.1103/PhysRevMaterials.3.111401}{Phys. Rev. Mater. \textbf{3}, 111401(R) (2019).}
\bibitem{Lin:prl19} L. F. Lin, Y. Zhang, A. Moreo, E. Dagotto, and S. Dong, \href{https://doi.org/10.1103/PhysRevLett.123.067601}{Phys. Rev. Lett. \textbf{123}, 067601 (2019).}
\bibitem{miao:pnas2020} T. Miao, L. Deng, W. Yang, J. Ni, C. Zheng, J. Etheridge, S. Wang, H. Liu, H. Lin, Y. Yu, Q. Shi, P. Cai, Y. Zhu, T. Yang, X. Zhang, X. Gao, C. Xi, M. Tian, X. Wu, H. Xiang, E. Dagotto, L. Yin, and J. Shen, \href{https://doi.org/10.1073/pnas.1920502117}{Proc. Natl. Acad. Sci. USA  \textbf{117}, 16226 (2020).}
\bibitem{zhu:prb102} Y. Zhu, B. Ye, Q. Li, H. Liu, T. Miao, L. Wu, L. Li, L.-F. Lin, Y. Zhu, Z. Zhang, Q. Shi, Y. Yang, K. Du, Y. Bai, Y. Yu, H. Guo, W. Wang, X. Xu, X. Wu, Z. Zhong, S. Dong, Y. Zhu, E. Dagotto, L. Yin, and J. Shen, \href{https://doi.org/10.1103/PhysRevB.102.235107}{Phys. Rev. B \textbf{102}, 235107 (2020).}
\bibitem{sen:2020} C. Sen and E. Dagotto, \href{https://doi.org/10.1103/PhysRevB.102.035126}{Phys. Rev. B \textbf{102}, 035126 (2020).}
\bibitem{Zhang:prb21} Y. Zhang, L. F. Lin, A. Moreo, G. Alvarez, and E. Dagotto, \href{https://doi.org/10.1103/PhysRevB.103.L121114}{Phys. Rev. B \textbf{103}, L121114 (2021).}
\bibitem{Lin:prl21} L. F. Lin, Y. Zhang, G. Alvarez, A. Moreo, and E. Dagotto, \href{https://doi.org/10.1103/PhysRevLett.127.077204}{Phys. Rev. Lett. \textbf{127}, 077204 (2021).}
\bibitem{mazza:2021} A. R. Mazza, E. Skoropata, J. Lapano, J. Zhang, Y. Sharma, B. L. Musico, V. Keppens, Z. Gai, M. J. Brahlek, A. Moreo, D. A. Gilbert, E. Dagotto, and T. Z. Ward, \href{https://doi.org/10.1103/PhysRevB.104.094204}{Phys. Rev. B \textbf{104}, 094204 (2021).}
\bibitem{zhang:2021} Y. Zhang, L.-F. Lin, G. Alvarez, A. Moreo, and E. Dagotto, \href{https://doi.org/10.1103/PhysRevB.104.125122}{Phys. Rev. B \textbf{104}, 125122 (2021).}
\bibitem{lin:2021} L.-F. Lin, N. Kaushal, C. Sen, A. D. Christianson, A. Moreo, and E. Dagotto,
\href{https://doi.org/10.1103/PhysRevB.103.184414}{Phys. Rev. B \textbf{103}, 184414 (2021).}
\bibitem{Kamihara:Jacs} Y. Kamihara, T. Watanabe, M. Hirano, and H. Hosono, \href{https://doi.org/10.1021/ja800073m}{J. Am. Chem. Soc. \textbf{130}, 3296 (2008).}
\bibitem{Dai:Np} P. C. Dai, J. P. Hu, and E. Dagotto, \href{https://doi.org/10.1038/nphys2438}{Nat. Phys. \textbf{8}, 709 (2012).}
\bibitem{Li:Nature} D. Li, K. Lee, B. Y. Wang, M. Osada, S. Crossley, H. R. Lee, Y. Cui, Yi, Y. Hikita, and H. Y. Hwang, \href{https://doi.org/10.1038/s41586-019-1496-5}{Nature \textbf{572}, 624 (2019).}
\bibitem{Zhang:prb20} Y. Zhang, L.-F. Lin, W. Hu, A. Moreo, S. Dong, and E. Dagotto, \href{https://doi.org/10.1103/PhysRevB.102.195117}{Phys. Rev. B \textbf{102}, 195117 (2020).}
\bibitem{Brink:jpcm} J. van den Brink and D. I. Khomskii, \href{https://doi.org/10.1088/0953-8984/20/43/434217}{J. Phys.: Condens. Matter {\bf 20}, 434217 (2008).}
\bibitem{Lin:prm17} L.-F. Lin, Q.-R. Xu, Y. Zhang, J.-J. Zhang, Y.-P. Liang, and S. Dong, \href{https://doi.org/10.1103/PhysRevMaterials.1.071401}{Phys. Rev. Materials \textbf{1}, 071401(R) (2017).}
\bibitem{Dong:nsr} S. Dong, H.-J. Xiang, and E. Dagotto, \href{https://doi.org/10.1093/nsr/nwz023}{Nat. Sci. Rev. {\bf 6}, 629 (2019).}
\bibitem{Zhang:prb20-2} Y. Zhang, L. F. Lin, A. Moreo, S. Dong, and E. Dagotto, \href{https://journals.aps.org/prb/abstract/10.1103/PhysRevB.101.144417}{Phys. Rev. B \textbf{101}, 144417 (2020).}
\bibitem{Tokura:science}Y. Tokura and N. Nagaosa, \href{https://doi.org/10.1126/science.288.5465.462}{Science {\bf 288}, 462 (2000)}.
\bibitem{Pandey:prb} B. Pandey, Y. Zhang, N. Kaushal, R. Soni, L.-F. Lin, W.-J. Hu, G. Alvarez, and E. Dagotto, \href{https://doi.org/10.1103/PhysRevB.103.045115}{Phys. Rev. B {\bf 103}, 045115 (2021).}
\bibitem{Lin:prm21} L.-F. Lin, N. Kaushal, Y. Zhang, A. Moreo, and E. Dagotto, \href{https://doi.org/10.1103/PhysRevMaterials.5.025001}{Phys. Rev. Mater. {\bf 5}, 025001 (2021)}.
\bibitem{DW} G. Gr{\"u}ner, {\it Density Waves in Solids} (Perseus, Cambridge, MA, 2000).
\bibitem{Zhang:prbcdw} Y. Zhang, L.-F. Lin,  A. Moreo, S. Dong, and E. Dagotto, \href{https://doi.org/10.1103/PhysRevB.101.174106}{Phys. Rev. B \textbf{101}, 174106 (2020).}
\bibitem{JH:prb1} E. {\c{S}}a{\c{s}}{\i}o{\u{g}}lu, C. Friedrich, and S. Bl\"ugel, \href{https://doi.org/10.1103/PhysRevB.83.121101}{Phys. Rev. B \textbf{83}, 121101(R) (2011);}
\bibitem{JH:prb2}L. Vaugier, H. Jiang, and S. Biermann, \href{https://doi.org/10.1103/PhysRevB.86.165105}{Phys. Rev. B \textbf{86}, 165105 (2012).}
\bibitem{SOC}W. Witczak-Krempa, G. Chen, Y. B. Kim, and L. Balents, \href{https://doi.org/10.1146/annurev-conmatphys-020911-125138}{Annu. Rev. Condens. Matter Phys. \textbf{5}, 57 (2014).}
\bibitem{Streltsovt:prb14} S. V. Streltsov and D. I. Khomskii, \href{https://doi.org/10.1103/PhysRevB.89.161112}{Phys. Rev. B {\bf 89}, 161112(R) (2014).}
\bibitem{Zhang:ossp} Y. Zhang, L. F. Lin, A. Moreo, and E. Dagotto, \href{https://doi.org/10.1103/PhysRevB.104.L060102}{Phys. Rev. B \textbf{104}, L060102 (2021).}
\bibitem{OSMP} L. de' Medici, S. R. Hassan, M. Capone, and X. Dai, \href{https://doi.org/10.1103/PhysRevLett.102.126401}{Phys. Rev. Lett. \textbf{102}, 126401 (2009).}
\bibitem{Patel:osmp} N. D. Patel, A. Nocera, G. Alvarez, A. Moreo, S. Johnston and E. Dagotto, \href{https://doi.org/10.1038/s42005-019-0155-3}{Comm. Phys. \textbf{2}, 64 (2019)}
\bibitem{Herbrych:osmp1} J. Herbrych, J. Heverhagen, N. D. Patel, G. Alvarez, M. Daghofer, A. Moreo, and E. Dagotto, \href{https://doi.org/10.1103/PhysRevLett.123.027203}{Phys. Rev. Lett. \textbf{123}, 027203 (2019).}
\bibitem{Herbrych:osmp2}  J. Herbrych, J. Heverhagen, N. D. Patel, G. Alvarez, M. Daghofer, A. Moreo, and E. Dagotto, \href{https://doi.org/10.1073/pnas.2001141117}{Proc. Natl. Acad. Sci. USA  \textbf{117}, 16226 (2020).}
\bibitem{Kim:prl08} B. J. Kim, Hosub Jin, S. J. Moon, J.-Y. Kim, B.-G. Park, C. S. Leem, Jaejun Yu, T. W. Noh, C. Kim, S.-J. Oh, J.-H. Park, V. Durairaj, G. Cao, and E. Rotenberg, \href{https://doi.org/10.1103/PhysRevLett.101.076402}{Phys. Rev. Lett. \textbf{101}, 076402 (2008).}
\bibitem{Jackeli:prl08} G. Jackeli and G. Khaliullin, \href{https://doi.org/10.1103/PhysRevLett.102.017205}{Phys. Rev. Lett. \textbf{102}, 017205 (2009).}
\bibitem{Lu:am20} C. Lu and J.-M. Liu \href{https://doi.org/10.1002/adma.201904508}{Adv. Mater. \textbf{32}, 1904508 (2020).}
\bibitem{Kitaev:aop} A. Kitaev, \href{https://doi.org/10.1016/j.aop.2005.10.005}{Annals of Physics  \textbf{321}, 2 (2006).}
\bibitem{Chaloupka:prl08} J. Chaloupka, G. Jackeli, and G. Khaliullin, \href{https://doi.org/10.1103/PhysRevLett.105.027204}{Phys. Rev. Lett. \textbf{105}, 027204 (2010).}
\bibitem{Singh:prl12} Y. Singh, S. Manni, J. Reuther, T. Berlijn, R. Thomale, W. Ku, S. Trebst, and P. Gegenwart, \href{https://doi.org/10.1103/PhysRevLett.108.127203}{Phys. Rev. Lett. \textbf{108}, 127203 (2012).}
\bibitem{Choi:prl12} S. K. Choi, R. Coldea, A. N. Kolmogorov, T. Lancaster, I. I. Mazin, S. J. Blundell, P. G. Radaelli, Yogesh Singh, P. Gegenwart, K. R. Choi, S.-W. Cheong, P. J. Baker, C. Stock, and J. Taylor, \href{https://doi.org/10.1103/PhysRevLett.108.127204}{Phys. Rev. Lett. \textbf{108}, 127204 (2012).}
\bibitem{Mazin:prl12} I. I. Mazin, Harald O. Jeschke, K. Foyevtsova, R. Valenti, and D. I. Khomskii, \href{https://doi.org/10.1103/PhysRevLett.109.197201}{Phys. Rev. Lett. \textbf{109}, 197201 (2012).}
\bibitem{Rau:prl14} J. G. Rau, Eric Kin-Ho Lee, and Hae-Young Kee, \href{https://doi.org/10.1103/PhysRevLett.112.077204}{Phys. Rev. Lett. \textbf{112}, 077204 (2014).}
\bibitem{Kim:prx20} J. Kim, J. Chaloupka, Y. Singh, J. W. Kim, B. J. Kim, D. Casa, A. Said, X. Huang, and T. Gog, \href{https://doi.org/10.1103/PhysRevX.10.021034}{Phys. Rev. X \textbf{10}, 021034 (2020).}
\bibitem{Bastien:prb18} G. Bastien, G. Garbarino, R. Yadav, F. J. Martinez-Casado, R. Beltr\'an Rodr\'{\i}guez, Q. Stahl, M. Kusch, S. P. Limandri, R. Ray, P. Lampen-Kelley, D. G. Mandrus, S. E. Nagler, M. Roslova, A. Isaeva, T. Doert, L. Hozoi, A. U. B. Wolter, B. B\"uchner, J. Geck, and J. van den Brink, \href{https://doi.org/10.1103/PhysRevB.97.241108}{Phys. Rev. B \textbf{97}, 241108(R) (2018).}
\bibitem{Li:prm19} G. Li, X. Chen, Y. Gan, F. Li, M. Yan, F. Ye, S. Pei, Y. Zhang, L. Wang, H. Su, J. Dai, Y. Chen, Y. Shi, X.W. Wang, L. Zhang, S. Wang, D. Yu, F. Ye, J.-W. Mei, and M. Huang, \href{https://doi.org/10.1103/PhysRevMaterials.3.023601}{Phys. Rev. Mater. \textbf{3}, 023601 (2019).}
\bibitem{Plumb:prb14} K. W. Plumb, J. P. Clancy, L. J. Sandilands, V. V. Shankar, Y. F. Hu, K. S. Burch, H.-Y. Kee, and Y.-J. Kim, \href{https://doi.org/10.1103/PhysRevB.90.041112}{Phys. Rev. B \textbf{90}, 041112(R) (2014).}
\bibitem{Johnson:prb15} R. D. Johnson, S. C. Williams, A. A. Haghighirad, J. Singleton, V. Zapf, P. Manuel, I. I. Mazin, Y. Li, H. O. Jeschke, R. Valent\'{\i}, and R. Coldea, \href{https://doi.org/10.1103/PhysRevB.92.235119}{Phys. Rev. B \textbf{92}, 235119 (2015).}
\bibitem{Koitzsch:prb16} A. Koitzsch, C. Habenicht, E. M\"uller, M. Knupfer, B. B\"uchner, H. C. Kandpal, J. van den Brink, D. Nowak, A. Isaeva, and Th. Doert, \href{https://doi.org/10.1103/PhysRevLett.117.126403}{Phys. Rev. Lett. \textbf{117}, 126403 (2016).}
\bibitem{Cao:prb16} H. B. Cao, A. Banerjee, J.-Q. Yan, C. A. Bridges, M. D. Lumsden, D. G. Mandrus, D. A. Tennant, B. C. Chakoumakos, and S. E. Nagler, \href{https://doi.org/10.1103/PhysRevB.93.134423}{Phys. Rev. B \textbf{93}, 134423 (2016).}
\bibitem{Banerjee:nm} A. Banerjee, C. A. Bridges, J.-Q. Yan, A. A. Aczel, L. Li, M. B. Stone, G. E. Granroth, M. D. Lumsden, Y. Yiu, J. Knolle, S. Bhattacharjee, D. L. Kovrizhin, R. Moessner, D. A. Tennant, D. G. Mandrus and S. E. Nagler, \href{https://doi.org/10.1038/nmat4604}{Nature Mater. \textbf{15}, 733 (2016).}
\bibitem{Kubota:prb15} Y. Kubota, H. Tanaka, T. Ono, Y. Narumi, and K. Kindo, \href{https://doi.org/10.1103/PhysRevB.91.094422}{Phys. Rev. B \textbf{91}, 094422 (2015).}
\bibitem{Glamazda:prb17} A. Glamazda, P. Lemmens, S.-H. Do, Y. S. Kwon, and K.-Y. Choi, \href{https://doi.org/10.1103/PhysRevB.95.174429}{Phys. Rev. B \textbf{95}, 174429 (2017).}
\bibitem{Kim:prb15} Kim, Heung-Sik and V., Vijay Shankar and Catuneanu, Andrei and Kee, Hae-Young, \href{https://doi.org/10.1103/PhysRevB.91.241110}{Phys. Rev. B \textbf{91}, 241110(R) (2015).}
\bibitem{Hou:prb17} Y. S. Hou, H. J. Xiang, and X. G. Gong, \href{https://doi.org/10.1103/PhysRevB.96.054410}{Phys. Rev. B \textbf{96}, 054410 (2017).}
\bibitem{Eichstaedt:prb19} C. Eichstaedt, Y. Zhang, P. Laurell, S. Okamoto, A. G. Eguiluz, and T. Berlijn, \href{https://doi.org/10.1103/PhysRevB.100.075110}{Phys. Rev. B \textbf{100}, 075110 (2019).}
\bibitem{Baek:prl17} S.-H. Baek, S.-H. Do, K.-Y. Choi, Y. S. Kwon, A. U. B. Wolter, S. Nishimoto, Jeroen van den Brink, and B. Buchner, \href{https://doi.org/10.1103/PhysRevLett.119.037201}{Phys. Rev. Lett. \textbf{119}, 037201 (2017).}
\bibitem{Zheng:prl17} J. Zheng, K. Ran, T. Li, J. Wang, P. Wang, B. Liu, Z.-X. Liu, B. Normand, J. Wen, and W. Yu, \href{https://doi.org/10.1103/PhysRevLett.119.227208}{Phys. Rev. Lett. \textbf{119}, 227208 (2017).}
\bibitem{Kasahara:prl17} Y. Kasahara, K. Sugii, T. Ohnishi, M. Shimozawa, M. Yamashita, N. Kurita, H. Tanaka, J. Nasu, Y. Motome, T. Shibauchi, and Y. Matsuda, \href{https://doi.org/10.1103/PhysRevLett.120.217205}{Phys. Rev. Lett. \textbf{120}, 217205 (2018).}
\bibitem{Do:np17} S.-H. Do, S.-Y. Park, J. Yoshitake, J. Nasu, Y. Motome, Y. S. Kwon, D. T. Adroja, D. J. Voneshen, K. Kim, T.-H. Jang, J.-H. Park, K.-Y. Choi and S. Ji , \href{https://doi.org/10.1038/nphys4264}{Nature Phys. \textbf{13}, 1079 (2017).}
\bibitem{Banerjee:science19} A. Banerjee, J. Yan, J. Knolle, C. A. Bridges, M. B. Stone,  M. D. Lumsden, D. G. Mandrus, D. A. Tennant, R. Moessner, and S. E. Nagler, \href{https://doi.org/10.1126/science.aah6015}{Science \textbf{356}, 1055 (2019).}
\bibitem{Banerjee:npjqm} A. Banerjee, P. Lampen-Kelley, J. Knolle, C. Balz, A. A. Aczel, B. Winn, Y. Liu, D. Pajerowski, J. Yan, C. A. Bridges, A. T. Savici, B. C. Chakoumakos, M. D. Lumsden, D. A. Tennant, R. Moessner, D. G. Mandrus and S. E. Nagler, \href{https://doi.org/10.1038/s41535-018-0079-2}{npj Quant Mater. \textbf{3}, 8 (2018).}
\bibitem{Kasahara:nature} Y. Kasahara, T. Ohnishi, Y. Mizukami, O. Tanaka, Sixiao Ma, K. Sugii, N. Kurita, H. Tanaka, J. Nasu, Y. Motome, T. Shibauchi and Y. Matsuda, \href{https://doi.org/10.1038/s41586-018-0274-0}{Nature \textbf{559}, 227 (2018).}
\bibitem{Ni:arXiv} D. Ni, X. Gui, K. M. Powderly, and R. J. Cava, \href{https://arxiv.org/abs/2108.12915}{arXiv:2108.12915v1}
\bibitem{Nawa:arXiv} K. Nawa, Y. Imai, Y. Yamaji, H. Fujihara, W. Yamada, R. Takahashi, T. Hiraoka, M. Hagihala, S. Torii, T. Aoyama, T. Ohashi, Y. Shimizu, H. Gotou, M.i Itoh, K. Ohgushi, T. J Sato, \href{https://arxiv.org/abs//2109.12864}{arXiv:/2109.12864v1}
\bibitem{Huang:prb16} C. Huang, J. Zhou, H. Wu, K. Deng, P. Jena, and E. Kan, \href{https://doi.org/10.1103/PhysRevB.95.045113}{Phys. Rev. B \textbf{95}, 045113 (2017).}
\bibitem{Ersan:jmmm19} F. Ersan, E. Vatansever, S. Sarikurt, Y. Y{\"u}ksel, Y. Kadioglu, H. D. Ozaydin, O. Akt{\"u}rk, {\"U}. Ak{\i}nc{\i}, and E. Akt{\"u}rk, \href{https://doi.org/10.1016/j.jmmm.2018.12.032}{J. Magn. Magn. Mater. \textbf{476}, 111 (2019).}
\bibitem{Kresse:Prb} G. Kresse and J. Hafner, \href{https://doi.org/10.1103/PhysRevB.47.558}{Phys. Rev. B \textbf{47}, 558 (1993).}
\bibitem{Kresse:Prb96} G.~Kresse and J.~Furthm\"{u}ller, \href{https://doi.org/10.1103/PhysRevB.54.11169}{Phys. Rev. B \textbf{54}, 11169 (1996).}
\bibitem{Blochl:Prb} P. E. Bl\"{o}chl, \href{https://doi.org/10.1103/PhysRevB.50.17953}{Phys. Rev. B \textbf{50}, 17953 (1994).}
\bibitem{Perdew:Prl} J. P. Perdew, K. Burke, and M. Ernzerhof, \href{https://doi.org/10.1103/PhysRevLett.77.3865}{Phys. Rev. Lett. \textbf{77}, 3865 (1996).}
\bibitem{Chaput:prb} L. Chaput, A. Togo, I. Tanaka, and G. Hug, \href{https://doi.org/10.1103/PhysRevB.84.094302}{Phys. Rev. B \textbf{84}, 094302 (2011).}
\bibitem{Togo:sm} A. Togo, I. Tanaka, \href{https://doi.org/10.1016/j.scriptamat.2015.07.021}{Scr. Mater. \textbf{108}, 1 (2015).}
\bibitem{Dudarev:prb} S. L. Dudarev, G. A. Botton, S. Y. Savrasov, C. J. Humphreys, and A. P. Sutton, \href{https://doi.org/10.1103/PhysRevB.57.1505}{Phys. Rev. B \textbf{57}, 1505 (1998).}
\bibitem{Kim:prb16} H.-S. Kim and H.-Y. Kee, \href{https://doi.org/10.1103/PhysRevB.93.155143}{Phys. Rev. B \textbf{93}, 155143 (2016).}
\bibitem{Eglitis:cry} R. I. Eglitis, J. Purans and R. Jia, \href{https://doi.org/10.3390/cryst11040455}{Crystals \textbf{11}, 455 (2021).}
\bibitem{Eglitis:sys} R. I. Eglitis, J. Purans, A. I. Popov and R. Jia, \href{ https://doi.org/10.3390/sym13101920}{Symmetry \textbf{13}, 1920 (2021).}
\bibitem{Momma:vesta} K. Momma and F. Izumi, \href{https://doi.org/10.1107/S0021889811038970}{J. Appl. Crystallogr. \textbf{44}, 1272 (2011).}
\bibitem{Perdew:Prl08} J. P. Perdew and A. Ruzsinszky, and G. I. Csonka and O. A. Vydrov and G. E. Scuseria and L. A. Constantin and X. Zhou and K. Burke, \href{https://doi.org/10.1103/PhysRevLett.100.136406}{Phys. Rev. Lett. \textbf{100}, 136406 (2008).}
\bibitem{Grimme:jcp} S. Grimme, J. Antony, S. Ehrlich, and S. Krieg, \href{https://doi.org/10.1063/1.3382344}{J. Chem. Phys. \textbf{132}, 154104 (2010).}
\bibitem{Grimme:jcc} S. Grimme, S. Ehrlich, and L. Goerigk, \href{https://doi.org/10.1002/jcc.21759}{J. Comp. Chem. \textbf{32}, 1456 (2011).}
\bibitem{Supplemental} For more results, see Supplemental Material at \href{http://link.aps.org/supplemental/10.1103/PhysRevB.xx/xxxxxx}{http://link.aps.org/supplemental/10.1103/PhysRevB.xx/xxxxxx.}
\bibitem{Savin:Angewandte} A. Savin, O. Jepsen, J. Flad, O.-K. Andersen, H. Preuss, and H. G. von Schnering, \href{https://doi.org/10.1002/anie.199201871}{Angew. Chem. Int. Ed. \textbf{32}, 187 (1992).}
\bibitem{Pollini:prb96} I. Pollini, \href{https://doi.org/10.1103/PhysRevB.53.12769}{Phys. Rev. B \textbf{53}, 12769 (1996).}
\bibitem{Binotto:pss} L. Binotto, I. Pollini, G. Spinolo, \href{https://doi.org/10.1002/pssb.2220440126}{Phys. Stat. Sol. \textbf{44}, 245 (1970).}
\bibitem{Sandilands:prb16} L. J. Sandilands, Y. Tian, A. A. Reijnders, H.-S. Kim, K. W. Plumb, Y.-J. Kim, H.-Y. Kee, and K. S. Burch, \href{https://doi.org/10.1103/PhysRevB.93.075144}{Phys. Rev. B \textbf{93}, 075144 (2016).}


\end{references}
\end{document}